\documentclass{llncs} % instead of \documentclass[11pt]{article}
\newtheorem{defn}{Definition} %[section]
\newtheorem{df}{Definition}%[defn]

\newtheorem{thm}{Theorem}
\newtheorem{lem}[defn]{Lemma}
\newtheorem{cor}{Corollary}
\newtheorem{rem}{Remark}
\newtheorem{con}{Conjecture}

\title{The length of a minimal synchronizing word and the \v{C}erny conjecture}

\date{11.07.2018}
\author{A.N. Trahtman\thanks{Email: avraham.trakhtman@gmail.com}
\institute{}
}

\begin{document}
\maketitle
\begin{abstract}
  A word $w$ of letters on edges of underlying graph $\Gamma$ of deterministic finite automaton (DFA) is called synchronizing if 
$w$ sends all states of the automaton to a unique state.

J. \v{C}erny discovered in 1964 a sequence of $n$-state complete DFA possessing a minimal synchronizing word of length $(n-1)^2$.

The hypothesis, well known today as the \v{C}erny conjecture, claims that it is also precise upper bound on the length of such a word for a complete DFA.
The hypothesis was formulated in 1966 by Starke.
The problem has motivated great and constantly growing number of investigations and generalizations. 

To prove the conjecture, we use algebra with non-standard operations on a special class of matrices (row monomial), 
induced by words in the alphabet of labels on edges.
These matrices generate a space with respect to the mentioned operation.

The proof is based on connection between length of words $u$ 
and dimension of the space generated by solutions $L_x$ 
of matrix equation $M_uL_x=M_s$ for synchronizing word $s$,
as well as on the relation between ranks of $M_u$ and $L_x$.

\end{abstract}

$\bf Keywords$: deterministic finite automaton, synchronizing word, \v{C}erny conjecture.
\section*{Introduction}

 The problem of synchronization of DFA is a natural one and various aspects of this problem have been touched in the literature.
Prehistory of the topic, the emergence of the term, the connections with the early coding theory,
first efforts to estimate the length of synchronizing word \cite{La}, \cite{Li}, different problems of synchronization one can find in surveys \cite{Ju}, \cite{KV}, \cite{Vo1}, \cite{Vo}.

Synchronization makes the behavior of an automaton resistant
against input errors since, after detection of an error,
a synchronizing word can reset the automaton back to its original
state, as if no error had occurred.
The synchronizing word limits the propagation of errors for a prefix code.

A problem with a long story is the estimation of the minimal length of synchronizing word.

 J. \v{C}erny in 1964 \cite{Ce} found the sequence of $n$-state complete DFA with shortest synchronizing word of length $(n-1)^2$ for an alphabet of size two.
  The hypothesis, well known today as the \v{C}erny's conjecture, claims that this lower bound on the length of the synchronizing word of aforementioned automaton is also the upper bound for the shortest synchronizing word of any $n$-state complete DFA:
\begin{con}
The deterministic complete $n$-state synchronizing automaton over alphabet $\Sigma$ has synchronizing word in $\Sigma$ of length at most $(n-1)^2$ \cite{Sta} (Starke, 1966).
  \end{con}

The problem can be reduced to automata with a strongly connected graph \cite{Ce}.
An attempt to prove this hypothesis is proposed below.

This famous conjecture is true for a lot of automata, but in general the problem still remains open although several hundreds of articles consider this problem from different points of view \cite{TB}.

Moreover, two conferences "Workshop on Synchronizing  Automata"
(Turku, 2004) and  "Around the  \v{C}erny conjecture"
(Wroclaw, 2008) were dedicated to this longstanding conjecture.
The problem is discussed in "Wikipedia" - the popular Internet
Encyclopedia and on many other sites.

As well as the Road Coloring problem \cite{AW}, \cite{Fi}, \cite{TP}, this simple-looking conjecture was arguably the most longstanding and famous open combinatorial problems in the theory of finite automata \cite{KV}, \cite{MS}, \cite{PS}, \cite{Sta}, \cite{St}, \cite{Vo}.

We consider a special class of matrices $M_u$ of mapping 
induced by words $u$ in the alphabet of letters on edges
of the underlying graph $\Gamma$. We call them matrices of word.   

The matrix $M_u$ of word $u$ belongs to the class of matrices with 
one unit in every row and rest zeros  (row monomial). 
Row monomial matrices also can be considered as  matrices of word of some suitable graph.

There are no examples of automata such that the length of
the shortest synchronizing word is greater than $(n-1)^2$.
Moreover, the examples of automata  with shortest synchronizing
word of length $(n-1)^2$ are infrequent.
After the sequence of \v{C}erny and the example of \v{C}erny, Piricka and Rosenauerova \cite{CPR} of 1971 for $|\Sigma|=2$,
the next such examples were found by Kari \cite {Ka} in 2001 for $n=6$ and $|\Sigma|=2$ and by Roman \cite {Ro} for $n=5$ and $|\Sigma|=3$ in 2004.

The package TESTAS \cite {TP}, \cite {Tt} studied all automata with strongly connected underlying graph of size $n \le 11$ for  $|\Sigma|=2$, of size $n \le 8$ for $|\Sigma| \le 3$ and of size $n \le 7$ for $|\Sigma| \le 4$ and found five new examples of DFA with shortest synchronizing word of length $(n-1)^2$ with $n\leq 4$.

 Don and Zantema present in \cite {DZ} an ingenious method of designing new automata from existing examples of size three and four and proved that for $n\geq 5$ the method does not work.
So there are up to isomorphism exactly 15 DFA for $n=3$ and exactly 12 DFA for $n=4$ with shortest synchronizing word of length $(n-1)^2$.
The authors of \cite {DZ} support the hypothesis from \cite{TS} that all automata with shortest synchronizing word of length $(n-1)^2$ are known, of course, with essential correction found by  themselves for $n=3,4$.

There are several reasons \cite{AGV}, \cite{BBP}, \cite{CA},  \cite {DZ}, \cite{TS} to believe that the length of the shortest synchronizing word
for remaining automata  with $n>4$ (except the sequence of \v{C}erny and two examples for $n=5, 6$) is essentially less and the gap grows with $n$.
For several classes of automata, one can find some estimations on the length in \cite{AGV}, \cite{CR}, \cite{Kr}, \cite{KKS}, \cite{Ta}.

Initially found upper bound for the minimal length of
synchronizing word was very big and 
has been consistently improved over the years by different authors.
The upper bound found by Frankl in 1982 \cite{Fr} is equal to
$(n^3-n)/6$.
The result was reformulated in terms of synchronization in \cite{Pin} and repeated independently in \cite{KRS}.

The cubic estimation of the bound exists since 1982. Attempts to improve Frankl's result were unsuccessful.

The considered deterministic automaton $A$ can be presented by a
complete underlying graph with edges labelled by letters of an alphabet.

Our work uses a special class of matrices $M_u$ of mapping induced by words $u$ in the alphabet of letters on edges
of the underlying graph (row monomial) with properties of corresponding space. 

We study the rational series $(S,u)$ (see \cite{BR}).
This approach for synchronizing  automata supposed first by B{\'e}al \cite{Be} proved to be fruitful \cite{BBP}, \cite{CA}, \cite{Co}.

We consider the equation $M_uL_x=M_s$ (\ref{ux}) for synchronizing word  $s$ and the space generated
by a sort of its row monomial solutions $L_x$.

A connection between the set of nonzero columns of matrix of word, subsets of states of automaton 
and our kind  $L_x$ of solutions  of (\ref{ux}) is revealed in Remarks.

There are no examples of automata such that the length of
the shortest synchronizing word is greater than $(n-1)^2$.
Moreover, the examples of automata  with shortest synchronizing
word of length $(n-1)^2$ are infrequent.
After the sequence of \v{C}erny and the example of \v{C}erny, Piricka and Rosenauerova \cite{CPR} of 1971 for $|\Sigma|=2$,
the next such examples were found by Kari \cite {Ka} in 2001 for $n=6$ and $|\Sigma|=2$ and by Roman \cite {Ro} for $n=5$ and $|\Sigma|=3$ in 2004.

The package TESTAS \cite {TP}, \cite {Tt} studied all automata with strongly connected underlying graph of size $n \le 11$ for  $|\Sigma|=2$, of size $n \le 8$ for $|\Sigma| \le 3$ and of size $n \le 7$ for $|\Sigma| \le 4$ and found five new examples of DFA with shortest synchronizing word of length $(n-1)^2$ with $n\leq 4$.

 Don and Zantema present in \cite {DZ} an ingenious method of designing new automata from existing examples of size three and four and proved that for $n\geq 5$ the method does not work.
So there are up to isomorphism exactly 15 DFA for $n=3$ and exactly 12 DFA for $n=4$ with shortest synchronizing word of length $(n-1)^2$.
The authors of \cite {DZ} support the hypothesis from \cite{TS} that all automata with shortest synchronizing word of length $(n-1)^2$ are known, of course, with essential correction found by  themselves for $n=3,4$.

There are several reasons \cite{AGV}, \cite{BBP}, \cite{CA},  \cite {DZ}, \cite{TS} to believe that the length of the shortest synchronizing word
for remaining automata  with $n>4$ (except the sequence of \v{C}erny and two examples for $n=5, 6$) is essentially less and the gap grows with $n$.
For several classes of automata, one can find some estimations on the length in \cite{AGV}, \cite{CR}, \cite{Kr}, \cite{KKS}, \cite{Ta}.

Initially found upper bound for the minimal length of
synchronizing word was very big and 
has been consistently improved over the years by different authors.
The upper bound found by Frankl in 1982 \cite{Fr} is equal to
$(n^3-n)/6$.
The result was reformulated in terms of synchronization in \cite{Pin} and repeated independently in \cite{KRS}.

The cubic estimation of the bound exists since 1982. Attempts to improve Frankl's result were unsuccessful.

The considered deterministic automaton $A$ can be presented by a
complete underlying graph with edges labelled by letters of an
alphabet.

Our work uses a special class of matrices $M_u$ of mapping induced by words $u$ in the alphabet of letters on edges
of the underlying graph (row monomial) with properties of corresponding space. 

We study the rational series $(S,u)$ (see \cite{BR}).
This approach for synchronizing  automata supposed first by B{\'e}al \cite{Be} proved to be fruitful \cite{BBP}, \cite{CA}, \cite{Co}.

We consider the equation $M_uL_x=M_s$ (\ref{ux}) for synchronizing word  $s$ and the space generated
by a sort of its row monomial solutions $L_x$.

A connection between the set of nonzero columns of matrix of word, subsets of states of automaton 
and our kind  $L_x$ of solutions  of (\ref{ux}) is revealed in Remarks.

Theorems  \ref{t}, \ref{t2} finish our attempt to prove the
\v{C}erny conjecture.
Theorem \ref{t4} and some corollaries contain certain consequences.

We consider the equation $M_uL_x=M_s$ (\ref{ux}) for synchronizing word  $s$ and the space generated
by a sort of its row monomial solutions $L_x$.

A connection between the set of nonzero columns of matrix of word, subsets of states of automaton 
and our kind  $L_x$ of solutions  of (\ref{ux}) is revealed in Remarks.

The ideas of the proof are illustrated on examples of automata with a maximal length of synchronizing word from \cite{Ka}, \cite{Ce}, \cite{Ro}.

 \section*{Preliminaries}
We consider a complete $n$-state DFA with
 strongly connected underlying graph $\Gamma$ and transition semigroup $S$
 over a fixed finite alphabet $\Sigma$ of labels on edges of  $\Gamma$ of an automaton $A$.
The trivial cases $n \leq 2$, $|\Sigma|=1$ and $|A \sigma|=1$ for
$\sigma \in\Sigma$ are excluded.

The restriction on strongly connected graphs is based on \cite{Ce}.
The states of the automaton $A$ are considered also as vertices of the graph $\Gamma$.

If there exists a path in an automaton from the state $\bf p$ to
the state $\bf q$ and the edges of the path are consecutively
labelled by $\sigma_1, ..., \sigma_k$, then for
$s=\sigma_1...\sigma_k \in \Sigma^+$ let us write ${\bf q}={\bf
p}s$.

Let $Px$ be the set of states ${\bf q}={\bf p}x$ for all ${\bf p}$
from the subset $P$ of states and $x \in \Sigma^+$.
Let $Ax$ denote the set $Px$ for the set $P$ of all states of the automaton.

 A word $s \in \Sigma^+ $ is called a {\it synchronizing (reset, magic, recurrent, homing, directable)} word of an automaton $A$ with underlying graph $\Gamma$ if $|As|=1$.
The word $s$ below denotes minimal synchronizing word such that for a state $\bf q$ $As=\bf q$.

The problem can be reduced to automata with a strongly connected graph \cite{Ce}.

The states of the automaton are enumerated with number one for the state $\bf q$.

 An automaton (and its underlying graph) possessing a synchronizing word is called {\it synchronizing}.

Let us consider a linear space generated by
$n\times n$-matrices $M$ with one unit in any row of the matrix and zeros everywhere else (row monomial).

We connect a mapping of the set of states of the automaton made by
a word $u$ with an $n\times n$-matrix $M_u$ such that for an element $m_{i,j} \in M_u$ takes place

\centerline{$m_{i,j}$= $\cases{1, &${\bf q}_i u ={\bf q}_j$; \cr 0, &otherwise.}$}

Any mapping of the set of states of the automaton  $A$  can be presented by some word $u$  
and by a corresponding matrix $M_u$.
For instance,

 \centerline{$M_u=\left(
\begin{array}{ccccccc}
  0 & 0 & 1 & . & . & . &  0 \\
  1 & 0 & 0 & . & . & . &  0 \\
  0 & 0 & 0 & . & . & . &  1 \\
  . & . & . & . & . & . &  . \\
  0 & 1 & 0 & . & . & . &  0 \\
  1 & 0 & 0 & . & . & . &  0 \\
\end{array}\right)
$}

 Let us call the matrix $M_u$ of the mapping induced by the word $u$, for brevity, the matrix of word $u$ and vice versa.

$M_uM_v=M_{uv}$ \cite{Be}.

The set of nonzero columns of $M_u$ (set of second indexes of its elements) of $M_u$ is denoted as $R(u)$.

The word $u$ of the matrix $M_u$ is called {\it irreducible} if
for every proper subword $v$ of $u$ $M_u \neq M_v$.

The minimal synchronizing word and all its subwords are irreducible.

Zero matrix is consideered as a matrix of empty word.

The subset of states $Au$ is denoted as $c_u$ with number of states $|c_u|$.
In $n$-vector $c_u$ the coordinate $j$ has unit if
the state $j \in c_u$ and zero in opposite case.

For linear algebra terminology and definitions, see \cite{Ln},
\cite{Ma}.

\section{Mappings induced by a word and subword}

\begin{rem} \label{r1}
The invertible matrix $M_a$ does not change the number of units of every column of $M_u$ in its image of
the product $M_aM_u$.

Every unit in the product $M_uM_a$ is the product of two units, first unit from nonzero column of $M_u$
and second unit from a row with one unit of $M_a$.

\end{rem}

\begin{rem} \label{r4}

The columns of the matrix $M_uM_a$ are obtained by permutation of columns $M_u$.
Some columns can be merged (units of columns are moved along
row to a common column) with $|R(ua)|<|R(u)|$.

The rows of the matrix $M_aM_u$ are obtained by permutation of rows of the matrix $M_u$.
Some of these rows may disappear and replaced by another rows of $M_u$.

\end{rem}

\begin{lem} \label{l1}

The number of nonzero columns $|R(b)|$ is equal to the rank of $M_b$.

\centerline{$|R(ua)| \leq |R(u)|$} and

\centerline{$R(au) \subseteq R(u)$.}

For invertible matrix $M_a$ $R(au)=R(u)$ and $|R(ua)|=|R(u)|$.

For the set of states of deterministic finite automaton $A$ and any words $u$ and $a$ $Aua \subseteq Aa$.

 Nonzero columns of $M_{ua}$ have units also in $M_a$.

\end{lem}

Proof.
The matrix $M_b$ has submatrix with nonzero determinant having only one unit in every row and in every nonzero column 
.
Therefore $|R(b)|$ is equal to the rank of $M_b$.

The matrix $M_a$ in the product $M_uM_a$ shifts column of
$M_u$ to columns of $M_uM_a$ without
changing the column itself by Remark \ref{r4} or merging.
some columns of $M_u$.

In view of possible merged columns, $|R(ua)|\leq |R(u)|$.

Some rows of $M_u$ can be replaced in $M_aM_u$ by another row and therefore some rows from $M_u$ may be changed,
but zero columns of $M_u$ remain in $M_aM_u$ (Remark 1).

Hence $R(au) \subseteq R(u)$ and $|R(ua)| \leq |R(u)|$.

For invertible matrix $M_a$ in view of existence $M_a^{-1}$ we have $|R(ua)|=|R(u)|$ and $R(au)= R(u)$.

From $R(ua) \subseteq R(a)$ follows $Aua \subseteq Aa$.

Nonzero columns of $M_{ua}$ have units also in $M_a$ in view of $R(ua) \subseteq R(a)$.

\begin{cor}  \label{c1a}
The invertible matrix $M_a$ keeps the number of units of any column of $M_u$ in corresponding column of the product
$M_aM_u$.
\end{cor}

\begin{cor}  \label{c1}
The matrix $M_s$ of word $s$ is synchronizing if and only
if $M_s$ has zeros in all columns except one and units in
the residuary column. 

All matrices of right subwords of $s$ also have at least one unit in this column.
\end{cor}

\begin{lem}\label{lam}
 Suppose that for row monomial matrices $M_i$ 
 and $M$ 
\begin{equation}
M =\sum_{i=1}^k\lambda_i M_i. \label{lm}
\end{equation}
with coefficients $\lambda$ from $Q$.

Then the sum $\sum^k_{i=1}\lambda_i =1$ and the sum $S_j$ of values in every row $j$ 
of the sum in (\ref{lm}) also is equal to one.

If $\sum^k_{i=1}\lambda_iM_i=0$  then $\sum_{i=1}^k \lambda_i=0$ and $S_j=0$ 
for every $j$ with $M_u=0$.

If the sum $\sum^k_{i=1}\lambda_i$ in every row is not unit
[zero] then $\sum_{i=1}^k\lambda_i M_i$
is not a row monomial matrix.
\end{lem}

Proof.
The nonzero matrices $M_i$ have $n$ cells with unit in the cell.
Therefore, the sum of values in all cells of the matrix $\lambda_i M_i$ is $n \lambda_i$.

For nonzero $M$ the sum is $n$. So one has in view of
$M =\sum_{i=1}^k\lambda_i M_i$

\centerline {$n=n\sum_{i=1}^k \lambda_i$, whence $1 =\sum_{i=1}^k \lambda_i$.}
Let us consider the row $j$ of matrix $M_j$ in (\ref{lm}) and let  $1_j$ be unit in the row $j$.
The sum of values in a row of the sum (\ref{lm}) is equal to unit in the row of $M$.
So $1 =\sum_{i=1}^k \lambda_i1_i=\sum_{i=1}^k \lambda_i$.

$\sum_{i=1}^k\lambda_i M_i=0$ implies $S_j=\sum_{i=1}^k \lambda_i1_i=\sum_{i=1}^k \lambda_i=0$ 
for  every row $j$.

If the matrix $M=\sum_{i=1}^k\lambda_i M_i$ is a matrix
of word or zero matrix then
$\sum^k_{i=1}\lambda_i \in \{0, 1\}$.
If $\sum^k_{i=1}\lambda_i\not\in \{0, 1\}$ or
the sum  in ${0, 1}$ is not the same in every row then we have opposite case and
the matrix does not belong  to the set of row monomial matrix.
\\
\\
The set of row monomial matrices is closed with respect 
to the considered operation and together with zero matrix generates a space.

\begin{lem}  \label {v3}
 The set $V$ of all $n\times k$-matrices of words 
(or $n\times n$-matrices with zeros in fixed $n-k$ columns for $k<n$) has $n(k-1)+1$ linear independent matrices.
 \end{lem}
\begin{proof}
Let us consider distinct $n\times k$-matrices of word with at most only one nonzero cell outside the last nonzero column $k$.

Let us begin from the matrices $V_{i,j}$ with unit in $(i,j)$ cell ($j<k$) and units in ($m,k$) cells for all $m$ except $i$.
The remaining cells contain zeros.
So we have $n-1$ units in the $k$-th column and only one unit in remaining $k-1$ columns of the matrix $V_{i,j}$.
Let the matrix $K$ have units in the $k$-th column and zeros in the other columns.
There are $n(k-1)$ matrices $V_{i,j}$. Together with $K$ they belong to the set $V$.
So we have $n(k-1)+1$ matrices. For instance,

\begin{picture}(0,40)
\end{picture}
$V_{1,1}=\left(
\begin{array}{cccccccc}
  1 & 0 & 0 & . & . & 0  \\
  0 & 0 & 0 & . & . & 1  \\
  0 & 0 & 0 & . & . & 1  \\
  . & . & . & . & . & .  \\
  0 & 0 & 0 & . & . & 1  \\
  0 & 0 & 0 & . & . & 1  \\
\end{array}
\right)$
\begin{picture}(4,40)
\end{picture}
$V_{3,2}=\left(
\begin{array}{cccccccc}
  0 & 0 & 0 & . & . & 1  \\
  0 & 0 & 0 & . & . & 1  \\
  0 & 1 & 0 & . & . & 0  \\
  . & . & . & . & . & .  \\
  0 & 0 & 0 & . & . & 1  \\
  0 & 0 & 0 & . & . & 1  \\
\end{array}
\right)$
\begin{picture}(4,40)
\end{picture}
$K=\left(
\begin{array}{cccccccc}
  0 & 0 & 0 & . & . & 1 \\
  0 & 0 & 0 & . & . & 1 \\
  0 & 0 & 0 & . & . & 1 \\
  . & . & . & . & . & . \\
  0 & 0 & 0 & . & . & 1 \\
  0 & 0 & 0 & . & . & 1 \\
\end{array}
\right)$

 The first step is to prove that the matrices $V_{i,j}$ and $K$ generate the space with the set $V$.
For arbitrary matrix $T$ of word from $V$ for every $t_{i,j} \neq 0$ and $j<k$,
let us consider the matrices $V_{i,j}$ with unit in the cell $(i,j)$ and the sum of them $\sum V_{i,j}=Z$.

The first $k-1$ columns of $T$ and $Z$ coincide.
   Hence in the first $k-1$ columns of the matrix $Z$ there is at most only one unit in any row.
 Therefore in the cell of $k$-th column of $Z$ one can find only value of $m$ or $m-1$.
The value of $m$ appears if there are only zeros
in other cells of the considered row. Therefore $\sum V_{i,j} - (m-1)K=T$.
Thus every matrix from the set $V$ is a span of $(k-1)n +1$ matrices from $V$.
It remains now to prove that the set of matrices $V_{i,j}$ and $K$ is a set of linear independent matrices.

If one excludes a certain matrix $V_{i,j}$ from the set of these matrices, then it is impossible to obtain a nonzero value in the cell $(i,j)$ and therefore to obtain the matrix $V_{i,j}$.
So the set of matrices $V_{i,j}$ is linear independent.
Every non-trivial linear combination of the matrices $V_{i,j}$ equal to a matrix of word has at least one nonzero element in the first $k-1$ columns.
Therefore, the matrix $K$ could not be obtained as a linear combination of the matrices $V_{i,j}$.
Consequently the set of matrices $V_{i,j}$ and $K$ forms a basis of the set $V$.
\end{proof}

\begin{cor}  \label {c2}
The set of all row monomial $n \times(n-1)$-matrices of words
(or $n\times n$-matrices with zeros in a fixed column)
has $(n-1)^2$ linear independent matrices.

The set of row monomial $n\times n$-matrices has at most
$n(n-1)+1$ linear independent matrices.

 \end{cor}
Proof. For $k=n-1$ it follows from $n(n-1-1)+1= (n-1)^2$.

\begin{cor}  \label {cp}
Suppose the vertex ${\bf  p} \not\in A \alpha$ and let words $u$ of matrices $M_u$ have the last letter $\alpha$.

Then there are at most $(n-1)^2$ linear independent matrices $M_u$.
 \end{cor}
Proof. All matrices $M_u$ have common zero column ${\bf p}$  because ${\bf  p} \not\in A \alpha$. 
So we have $n\times n$-matrices with zeros in a fixed column
and due to Corollary \ref{c2}
there are at most $(n-1)^2$ linear independent matrices $M_u$.

 \begin{cor}  \label {c3}
There are at most $n(n-1)+1$ linear independent matrices of words in the set of $n\times n$-matrices.
 \end{cor}

\begin{lem} \label{l3} {\bf Distributivity from left.}

For every words $b$ and $x_i$

\centerline{$M_b\sum\tau_iM_{x_i}=\sum\tau_iM_bM_{x_i}$.}
If $\sum\tau_iM_{x_i}$ is a matrix of word then 
also $\sum\tau_iM_bM_{x_i}$ is a matrix of word. 

\end{lem}

\begin{proof}
The matrix $M_b$ shifts rows of every $M_{x_i}$ and of the sum of them in the same way according to Remark \ref{r4}.
$M_b$ removes common row of them and replace also by common row
(Remark \ref{r4}).

Therefore the matrices $M_bM_{x_i}$ and the sum
$\sum\tau_iM_bM_{x_i}$ has the origin rows with one unit from
$M_{x_i}$ and from its linear combination $\sum\tau_iM_{x_i}$, maybe in another order.

If the matrix $\sum\tau_iM_{x_i}$ is a matrix of word then 
also the matrix $M_b\sum\tau_iM_{x_i}= \sum\tau_iM_bM_{x_i}$ is a matrix of word with unit in every row.

\end{proof}
Let us notice that from right it is sometimes  wrong.

\section{Rational series}
The section follows ideas and definitions from \cite{BR} and \cite{Be}.
We recall that a formal power series with coefficients in a field $K$ and variables in $\Sigma$ is a mapping of the free monoid $\Sigma^*$ into $K$ \cite{BR}, \cite{CA}.

We consider an $n$-state automaton $A$. Let $P$ denote the subset
of states of the automaton with the characteristic column vector
$P^t$ of $P$ of length $n$ having units in coordinates corresponding to the states of $P$ and zeros everywhere else. Let $C$ be a row of units of length $n$.
 Following \cite{Be}, we denote by $S$ the {\it rational series} depending on the set $P$ defined by:
\begin{equation}
(S,u) = C M_uP^t-C P^t= C(M_u-E)P^t. \label{ser}
\end{equation}

\begin{rem} \label{r3}
Let $S$ be a rational series depending on the set $P$.

If the cell $i$ in $P^t$ has zero
then $(S,u)$ does not depend on column $i$ of $M_u$.
If this cell $i$ has unit then the column $i$ of $M_u$ with $k$ units from (\ref{ser}) added to $(S,u)$ the value of $k-1$. 

For $k$ units in the column $q$ of $M_u$ and $P=\{\bf q\}$
$(S,u)=k-1$.
\end{rem}

\begin{lem} \label{v4}
Let $S$ be a rational series depending on the
set $P$ of an automaton $A$.
Let $M_u=\sum_{j=1}^k\lambda_j M_{u_j}$.
Then $(S,u) =\sum_{j=1}^k\lambda_j (S,u_j)$.

If $(S,u_j)=i$ for every $j$ then also $(S,u)=i$.
\end{lem}
\begin{proof}
One has in view of (\ref{ser})

\centerline {$(S,u)= C(\sum^k_{j=1}\lambda_jM_{u_j}-E)P^t$}
where $C$ is a row of units and $P^t$ is a characteristic
column of units and zeros.

Due to Lemma \ref{lam}

$\sum^k_{j=1}\lambda_jM_{u_j}-E=\sum^k_{j=1}\lambda_jM_{u_j}-\sum^k_{j=1}\lambda_j E =
\sum^k_{j=1} \lambda_j(M_{u_j}-E)$.
So
$(S,u)=C(M_u-E)P^t = C(\sum^k_{j=1}\lambda_j M_{u_j}-E)P^t =
C(\sum^k_{j=1}\lambda_j (M_{u_j}-E))P^t=
 \sum^k_{j=1}\lambda_jC(M_{u_j}-E)P^t=
\sum^k_{j=1}\lambda_j(S, u_j)$.

Thus $(S,u) =\sum_{j=1}^k\lambda_j (S,u_j)$.

If $\forall j$ $(S,u_j)=i$, then 
$(S,u)= \sum^k_{j=1}\lambda_j i=i\sum^k_{j=1}\lambda_j=i$ by Lemma \ref{lam}.
\end{proof}

From Lemma \ref{v4} follows
\begin{cor} \label{c4}
Let $S$ be a rational series depending on the
set $P$ of an automaton $A$.

The matrices $M_{u}$ with constant $(S,u)=i$ generate a space
$V$ such that for every nonzero matrix $M_t \in V$ of word $t$ $(S,t)=i$.
\end{cor}

\begin{cor}\label{c5}
Let $S$ be a rational series depending on the set $P$ of size one of $n$-state automaton.

Then the set $V$ of matrices $M_{u}$ with two fixed nonzero columns and fixed nonnegative $(S,u) <n-1$ has at most $n$ linear independent matrices.
\end{cor}
\begin{proof}

By lemma \ref{v3} for $k=2$ there are at most $n+1$ linear independent matrices.
There is a matrix $M_w$ in a space for $k=2$ with one nonzero column and $(S,w)\neq (S,u)$.
Therefore fixed $(S,u)<n-1$ excludes the matrix $M_w$ from space generated by $V$.
\end{proof}

\subsection{The equivalence for the state $\bf q$, rational series $S_q$  }

\begin{df}
Two matrices $M_u$ and $M_v$ of word are called {\emph q-equivalent}
if the columns of the state $\bf q$ of both matrices are equal.
We denote it as

\centerline{$M_u \sim_q M_v$.} 

If the set of cells with units in
the column $\bf q$ of the matrix $M_v$ is a subset of the
analogous set of the matrix $M_u$ then we write

\centerline{$M_v \sqsubseteq_q M_u$}

Let rational series $S_q$ depend only on the set $P =\{\bf q\}$  for the state $q$  of number one.

\end{df}

Of course, for $As=\bf q$ and rational series depending on the set $P=\{\bf q\}$ for matrices 
of words in the alphabet $\Sigma$.

\begin{rem} \label{r6}

 The matrix $M_u$ has $(S_q,u)+1$ units in the column $q$.

$(S_q,u)=(S_q,v)$ if $M_u \sim_q M_v$ and 

$(S_q,v)\leq (S_q,u)$ if $M_v \sqsubseteq_q M_u$

\end{rem}

\begin{lem}\label{v20}
 
For matrices $M_{\alpha}$, $M_u$, $M_v$ of words $u$, $v$

\centerline{$M_u \sim_q M_v \to M_{\alpha u}M_u=M_{\alpha}M_u \sim_q M_{\alpha}M_v=M_{\alpha u}$,}

\centerline{$M_v \sqsubseteq_q M_u \to M_{\alpha}M_v \sqsubseteq_q M_{\alpha}M_u$.}

\end{lem}
\begin{proof}
Suppose $M_u \sim_q M_v$ and element $a_{i,r}=1$ in $M_{\alpha}$.
For an element $u_{r,q}$ in the column $q$ of $M_u$ and $t_{i,q} \in M_t=M_{\alpha}M_u$

\centerline{$t_{i,q}=\sum_{m=1}^n a_{i,m}u_{m,q}=a_{i,r}u_{r,q}$}
because $a_{i,m}=0$ for $m\neq r$ in the matrix $M_{\alpha}$ of word $a$ (Remark \ref{r1}).
Analogously, in the matrix $Z=M_{av}$ $z_{i,q}=a_{i,r}v_{r,q}$.

Therefore $z_{i,q}=a_{i,r}v_{r,q}=a_{i,r}u_{r,q}=t_{i,q}$ because $v_{r,q}=u_{r,q}$
for every cell $(i,q)$ of the column $q$ of $M_u$ and $M_v$.

Thus matrices $M_{au}$ and $M_{av}$ have common columns $\bf q$.
So $M_u \sim_q M_v$ implies $M_{au} \sim_q  M_{av}$.

Suppose now $M_v \sqsubseteq_q M_u$. For the matrix $T=M_{av}$
with $t_{i,q}=1$ one has $t_{i,q}=a_{i,r}v_{r,q}=1$ 
for some $v_{r,q}=1$ and $a_{i,r}=1$ as well as before. 
From $v_{r,q}=1$
and $M_v \sqsubseteq_q M_u$ follows $1=v_{r,q}=u_{r,q}$. 
So for the matrix $W=M_{au}$ one has $w_{i,q}=a_{i,r}u_{r,q}=1$, whence $t_{i,q}=1$ implies $w_{i,q}=1$ for every $i$.

Thus $M_v \sqsubseteq_q M_u$ implies $M_{\alpha}M_v \sqsubseteq_q M_{\alpha}M_u$.
\end{proof}

From Lemma \ref{v20} follow

\begin{cor}\label{c12}
For $As=\bf q$ and $M_u \sim_q M_v$

\centerline{$M_s \sim_q M_tM_v \to M_s=M_tM_u=M_tM_v.$}
For  $M_v \sqsubseteq_q M_u$

\centerline{$M_s = M_tM_v  \to M_s=M_tM_u.$}

$M_vL_u \sim_q L_{vu}$ for any words $u$, $v$.

\end{cor}

In the following example  $V_1 \sim_q V_2$ for the first column
$\bf q$, $M_s=M_{\alpha}V_1 =M_{\alpha}V_2$.
\\
\\
$M_{\alpha}=\left(
\begin{array}{cccccccc}
  0 & 1 & 0 & 0 & 0 \\
  0 & 1 & 0 & 0 & 0 \\
  0 & 1 & 0 & 0 & 0 \\
  0 & 0 & 1 & 0 & 0 \\
  0 & 0 & 1 & 0 & 0 \\
\end{array}
\right)$
$V_1=\left(
\begin{array}{cccccccc}
  0 & 0 & 0 & 1 & 0 \\
  1 & 0 & 0 & 0 & 0 \\
  1 & 0 & 0 & 0 & 0 \\
  0 & 0 & 1 & 0 & 0 \\
  0 & 0 & 0 & 0 & 1 \\
\end{array}
\right)$
$V_2=\left(
\begin{array}{cccccccc}
  0 & 0 & 1 & 0 & 0 \\
  1 & 0 & 0 & 0 & 0 \\
  1 & 0 & 0 & 0 & 0 \\
  0 & 0 & 0 & 0 & 1 \\
  0 & 0 & 0 & 1 & 0 \\
\end{array}
\right)$
$M_{\alpha}V_i=\left(
\begin{array}{cccccccc}
  1 & 0 & 0 & 0 & 0 \\
  1 & 0 & 0 & 0 & 0 \\
  1 & 0 & 0 & 0 & 0 \\
  1 & 0 & 0 & 0 & 0 \\
  1 & 0 & 0 & 0 & 0 \\
\end{array}
\right)$

\section{The equation with unknown $L_x$}

Rational series $S_q$ depends on the state $\bf q$.
$A s = \bf q$ for irreducible synchronizing word $s$.

The solution $L_x$ of the equation
\begin{equation}
M_uL_x=M_s \label{ux}
\end{equation}
for synchronizing matrix $M_s$ and arbitrary $M_u$
must have units in the column of the state $\bf q$.

\begin{lem}  \label{l5}
Every equation $M_uL_x=M_s$ (\ref{ux}) has a solutions $L_x$ with $n>(S_q,x) \geq 0$.

 $|R(u)|-1=(S_q,x)$ for $L_x$ with minimal $(S_q,x)$ (a minimal solution),
every matrix $L_y$ satisfies the equation (\ref{ux}) iff
$L_x\sqsubseteq_q L_y$.

There exists one-to-one correspondence between units in the column $q$ of minimal solution $L_x$ 
and the set $c_u$ of states.
Every set $c\supset c_u$ is presented by units of column $q$ of some not minimal solution $L_y$ of the equation.
\end{lem}

\begin{proof}
The matrix $M_s$ of rank one has nonzero column of the state $\bf q$.
For every nonzero column $j$ of $M_u$ with elements $u_{i,j}=1$ and $s_{i,q}=1$ in the matrix $M_s$ let the cell $(j,q)$ 
have unit in the matrix $L_x$.
So the unit in the column $q$ of matrix $M_s$ is a product of

every unit from the column $j$ of $M_u$ 
and unit in the sell $j$ of column $q$ of $L_x$.

The set $R(u)$ of nonzero columns of $M_u$ corresponds the set of cells of the column $q$ with unit of $L_x$.

Therefore by Remark \ref{r3} for rational series $S$ that depends on the state $\bf q$ the minimal solution 
$L_x$ has in the column $q$
$(S_q,x)+1$ units, whence $(S_q,x)=|R(u)|-1$.

So to the column $q$ of every solution belong at least $(S_q,x)+1$ units.
The remaining units of the solution $L_x$ belong to the next
columns, one unit in a row.
The remaining cells obtain zero.

Lastly every solution $L_x$ is a matrix of word.

Zeros in the column $q$ of minimal $L_x$ correspond zero columns of $M_u$. 
Therefore for matrix $L_y$ such that $L_x \sqsubseteq_q L_y$
we have $M_uL_y=M_s$.
On the other hand, every solution $L_y$ must have units in cells of column $q$ that correspond nonzero columns of $M_u$.

Thus $L_x$ has minimal $(S_q,x)$ and the equality $M_uL_x=M_uL_y=M_s$ is
equivalent to $L_x \sqsubseteq_q L_y$.

The matrix $M_u$ has set $R(u)$ of nonzero columns and maps the automaton on the set $c_u$ of states and on the set of units in the column $q$ of minimal $L_x$.
Units in the column $q$ of $L_y$ correspond some set of states $c\supset c_u$.

\end{proof}

Lemma \ref{l5} explains the following

\begin{rem} \label{r7}

Every permutation and shift of $m$ nonzero columns $M_u$
induces corresponding  permutation of the set of $m$ units in
the column $q$ of minimal solution $L_x$ of (\ref{ux}), and vice versa.

\end{rem}

\begin{df} \label{inv}
Let us call the matrix $M_a^-$ of word 
{\sf left generalized inverse} matrix of the matrix $M_a$ of a word $a$ if for precisely one element $a_{i,j}=1$ of every nonzero column $j$ of $M_a$ the cell $(j,i)$ of $M_a^-$ has unit.

 If in $M_a^-$ still are zero rows then unit is added arbitrarily in such row of the matrix $M_a^-$ of word.
\end{df}
For invertible matrix $M_a$ we have
$M_a^-=M_a^{-1}$, for singular $M_a$ there are some
generalized inverse matrices, including invertible.

\subsection{Right pseudoinverse matrices}

\begin{defn} \label{inv}

Let us call the matrix $M_a^-$ of word $a^-$
{\sf right pseudoinverse} matrix of the matrix $M_a$
of word $a$ if for precisely one element $a_{i,j}=1$ of
every nonzero column $j$ of $M_a$ the cell $(j,i)$ of
$M_a^-$ has unit.

In still zero rows of $M_a^-$ is added one unit arbitrarily in every such row.
Zeros fill rest of cells. So it is a matrix of word.

\end{defn}
For instance,
\\
\\
\noindent$M_a=\left(
\begin{array}{cccccccc}
  0 & 1 & 0 & 0 & 0 \\
  0 & 1 & 0 & 0 & 0 \\
  0 & 0 & 0 & 0 & 1 \\
  0 & 0 & 1 & 0 & 0 \\
  0 & 0 & 1 & 0 & 0 \\
\end{array}
\right)$
$M_a^-=\left(
\begin{array}{cccccccc}
  0 & 1 & 0 & 0 & 0 \\
  \d{1} & 0 & 0 & 0 & 0 \\
  0 & 0 & 0 & 0 & \d{1} \\
  0 & 0 & 0 & 1 & 0 \\
  0 & 0 & \d{1} & 0 & 0 \\
\end{array}
\right)$
$M_a^-=\left(
\begin{array}{cccccccc}
  1 & 0 & 0 & 0 & 0 \\
  0 & \d{1} & 0 & 0 & 0 \\
  0 & 0 & 0 & \d{1} & 0 \\
  0 & 0 & 0 & 1 & 0 \\
  0 & 0 & \d{1} & 0 & 0 \\
\end{array}
\right)$

\begin{rem} \label{ri}
For invertible matrix $M_a$ (with $|R(a)|=n$) we have
a special case $M_a^-=M_a^{-1}$, for singular $M_a$ there are some pseudoinverse matrices, even some invertible.

The product $M_aM_a^-$ does not depend on arbitrary adding of units in rows of $M_a^-$ corresponding
zero columns of $M_a$ in view of Remark \ref{r1}, because the nonzero product needs at least one nonzero cell in corresponding column of $M_a$.
\end{rem}

\begin{rem} \label{rm}
Some matrix $M_s^-$ with $As=q$ defines set of paths of $s$ from the state $q$ in opposite direction to every state.

Some matrix $M_a^-$ defines several paths of $a$ from the
state $q$ in opposite direction to the set of corresponding
states $c_a$.

 By Definition \ref{inv}  and Lemma \ref{l3} for  $M_b^-$ and $ M_a^-=\sum \lambda_iM_{a_i}^-$ one has

\centerline{$M_b^-M_a^-=M_{ab}^-$ and $M_b^-M_a^-=M_b^-\sum \lambda_i M_{a_i}^-=\sum \lambda_iM_b^-M_{a_i}^-$.}

\end{rem}

\begin{lem} \label{l7}
For every equation $M_uL_x=M_s$ and every letter $\beta$ the equation

\begin{equation}
 M_{u\beta}L_y=M_s      \label{yx}
\end{equation}
has solution $L_y$. For minimal solutions $L_x$ of (\ref{ux}) and $L_y$ one has $(S_q,y) \leq (S_q,x)$.  
$R(y)\supseteq R(x)$ is possible for some $L_y$. 

For every solution $L_x$ of equation (\ref{ux}) and
suitable $M_{\beta}^-$, even invertible,

\centerline{$M_s=M_uM_{\beta}M_{\beta}^-L_x$}
for solution $M_{\beta}^-L_x$ of the equation (\ref{yx}).

Let $|R(u)|=|R(u\beta)|$.
Then $(S_q,y)=(S_q,x)$ for minimal solutions
$L_y$, $L_x$ and  maximal ranks $|R(y)|=|R(x)|$ for invertible $M_{\beta}^-$.
$R(y)=R(x)$ for invertible $M_{\beta}^-$ anyway.

For $|R(u)|\neq |R(u\beta)|$ and singular $M_{\beta^-}$
there exists solution $L_y$ of the equation
$M_uM_{\beta}L_y=M_{u\beta}L_y=M_s$
such that $(S_q,y)<(S_q,x)$ for minimal solutions and
$|R(y)|>|R(x)|$ for maximal ranks.
Thus for some $L_y$ one has $R(x) \subset R(y)$ and
$|R(y)|=|R(x)|+|R(u)|-|R(u\beta)|$.
 $|R(u)|+|R(x)|=n+1$ for last considered word $u$ and corresponding minimal solution $L_x$.

  \end{lem}

\begin{proof}
The equality in (\ref{yx}) is correct for some $L_y$.
By Lemma \ref{l1} $|R(u)|\geq |R(u\beta)|$.
Therefore by Corollary \ref{c7}
$(S_q,y) \leq (S_q,x)$ for minimal solutions $L_x$ and $L_y$.
Hence in view of arbitrary placing $n-(S_q,y)$ units
in $L_y$ outside column $q$ (Lemma \ref{l5}),
$R(x) \subseteq R(y)$ is possible for some minimal $L_y$.

The matrix $M_{\beta}^-$ returns the set of nonzero columns
from $R(u\beta)$ to $R(u)$
(or to its part) in view of Definition \ref{inv}.

Arbitrary placing of units in some rows of $M_{\beta}^-$ does not change the product $M_{\beta}M_{\beta}^-$
by Remark \ref{ri}.
Therefore $|R(u)M_{\beta}M_{\beta}^-|\leq |R(u)|$,
whence $(S_q,x)\geq(S_q,y)$.
Hence the equality in

\centerline{$M_uM_{\beta}M_{\beta}^-L_x=M_{u\beta}M_{\beta}^-L_x=M_{u\beta}L_y=M_s$}
is correct for some $L_y=M_{\beta}^-L_x$ with
$R(x)\subseteq R(y)$ and free placing only of
$(S_q,x)-(S_q,y)$ units in $L_y$ (see Lemma \ref{l5}).

In the case $|R(u)|=|R(u\beta)|$ the matrix $M_{\beta}$
does not merge some columns of $M_u$ and by Lemma \ref{l5}
$(S_q,y)=(S_q,x)$ for minimal solutions $L_y$ and $L_x$.
So $R(y)=R(x)$ in view of $R(x)\subseteq R(y)$ and
Lemma \ref{l5} for invertible matrix $M_{\beta}^-$.

From $|R(u)|\neq |R(u\beta)|$ due to Lemma \ref{l1} follows
$|R(u\beta)|<|R(u)|$, whence for some solution $L_y$
of the equation $M_uM_{\beta}L_y=M_s$
$(S_q,y)<(S_q,x)$ for both such minimal solutions by Lemma
\ref{l5}.

 After filling by units nonzero columns of $R(x)$ by units from $R(y)$,
$R(x)$ can be extended by new columns using arbitrary addition
of $|R(y)|-|R(x)|$ units and $R(y)\supset R(x)$.

The possible equalities $|R(x)|=n-(S_q,x)$ and
 $|R(u)|-1=(S_q,x)$ (Lemma \ref{l5}) imply
 for minimal $L_y$ and $L_x$ and maximal ranks
$|R(u)|+|R(x)|=n+1$,
$|R(y)|-|R(x)|=(S_q,x)-(S_q,y)$ and
$|R(y)|=n-(S_q,y)=|R(x)|+(S_q,x)-(S_q,y)=
|R(x)|+|R(u)|-|R(u\beta)|$.

\end{proof}

\begin{cor}  \label{c7}
Let $M_uL_x=M_s$ (\ref{ux}) and $|a|<n$ for words $a$.

A set of $m\leq n$ linear independent matrices $L_y=M_a^-L_x$ 
with $(S_q,y)=(S_q,x)$ can be created by help of invertible generalized inverse matrix $M_a^-$.

\end{cor}

From Lemma \ref{l7} follows

\begin{cor}  \label{c9}
A set of linear independent solutions $L_x$ of
(\ref{ux}) with constant $(S_q,x)$
and fixed $R(x)$
 can be expanded by help of invertible matrices
$M_{\beta}^-$   (and words of them)
with the same $(S_q,x)$ and common set $R(x)$.
\end{cor}
Proof.
The invertible matrix $M_{\beta}^-$ does not change $(S_q,x)$ and $R(x)$ of matrix $L_x$ in the equation (\ref{yx}) by
Corollary \ref{c1a}.

Anyway we have a space generated by row monomial matrices with one unit in every row and with rest of zeros.

\begin{rem} \label{r9}
Not minimal solutions $L_y$ of (\ref{yx}) with
$(S_q,y)>(S_q,x)$ and $R(y) \subset R(x)$ also are useful sometimes for extending subspace $V_k$ of greater
$(S_q,y)=n-k$.
Arbitrary placing of units in $L_y$ is preferable in nonzero columns of matrices of $V_k$.
\end{rem}

\begin{lem} \label{v8}

Let the space $W$ be generated by matrix $M_s$ of minimal synchronizing word $s$
and solutions $L_x$ of the equation $M_uL_x=M_s$ (\ref{ux}) with $(S_q,x)>0$ 
of words $u$.   All generators $W$  have common zero column.

Then there exist a word $u$ in equation (\ref{ux}) such that the solution $L_x \in W$ 
and a letter $\beta$ 
 such that  the solution $L_y \not\in W$ of the equation $M_{u\beta}L_y=M_s$. 
 
\end{lem}

 Proof. Assume the contrary: for every word $u$ of (\ref{ux}) with solution $L_x \in W$ 
and every letter $\beta$ the equation $M_{u\beta}L_y=M_s$ has every solution $L_y\in W$.
 
The space  $W$  is not changed by the assumption and has the same basis. 
The matrices in $W$ have the same zero column. 
 
The solution  $L_y$ is a linear combination of matrices from basis of $W$ and therefore  $L_y$ 
can replace one matrix from this linear combination thereby creating a new basis of $W$ with $L_y$. 

By assumption, every solution  $L_y$ of equation $M_{v}L_y=M_s$ 
belongs to $W$ for the word  $v=u\beta$ with $|v|\leq |u|+1$.

Let us consider the word $u\beta\delta=v\delta$  of length $|u|+2=|v|+1$ for arbitrary letter $\delta$ 
and solution  $L_z$  of  equation $M_{v\delta}L_z=M_vM_{\delta}L_z= M_s$ 
 for the word $v\delta$.

By assumption, the solution  $L_z$ of equation $M_{v\delta}L_z=M_s$ 
also belongs to $W$.

  By induction it is true for the word  $u$ from (\ref{ux}) and $v$ of every length, whence every solution $L_x$   
has the common zero column. $L_x$ with $(S_q,x) > 0$ 
also belongs to some basis of $W$ as a solution of (\ref{ux}).

The considered automaton is synchronizing, therefore for every word $u$ exists a synchronizing 
continuation $v$ of $u$ with $ |R (uv)| =1$ of synchronizing matrix $M_{uv}$. 
Now  $(S_q,x)=0$ by Lemma \ref{l5} in spite of $(S_q,x)>0$ above.
                        
Therefore the existence of common zero column in all solutions $L_x$ of (\ref{ux}) 
contradicts the emergence of minimal solution $L_x$ with $(S_q,x) = 0$
on some step of expanding of words $u$.

\section{The sequence of spaces of solutions $L_x$ for words $u$ of growing length}

We are going to define a sequence of spaces $W_j$ which is 
 is generated by $M_s$ of word $s$ and $j$ 
linear independent solutions $L_x$  of equations
 $M_uL_x=M_s$ (\ref{ux}) with $|u|\leq j$.

The space $W_0$, in particular, is generated by minimal synchronizing matrix $M_s$,
a trivial solution of every equation (\ref{ux}).
$\dim(W_0)=1$.
The matrix $M_s$ and the minimal solution $L_x$  of equation $M_{\alpha}L_x=M_s$
for the left letter $\alpha$ of minimal $s$ generate the subspace $W_1$. 
$\dim(W_1)=2$.

We consider for every $W_j$ the set of solutions $L_x$ of equation (\ref{ux}) for $|u|\leq j+1$.
We choose a solution $L_x \not\in W_j$ for minimal such
$|u|$ following Lemmas \ref{l5}, \ref{l7}.
The existence of such $L_x$ is studied in Lemma \ref{v8}.
Then $L_x$ is added to the space $W_j$ turning it into the space $W_{j+1}$ with
corresponding growth of $j$.

The solutions $L_x$ of equations $M_uL_x=M_s$ with fixed $(S_q,x)=n-i$ generate subspace $V_i\subseteq W_j$ with
the same $(S_q,x)$ by Corollary \ref{c1a}.
$V_i$ can be extended by help of invertible matrices
$M_{\beta}^-$ of letters with keeping by Corollary \ref{c1a} the same rank $|R(x)|$.
In view of Lemma \ref{l7} the set of nonzero columns in
matrices $V_i$ does not changed because $M_{\beta}^-$
is invertible.

We can extend $V_i$ following Corollary \ref{c9} or reduce $(S_q,x)=i$.

The space $W_j$ is created by generators $L_x$ of subspaces $V_i$ by decrease of $i$ from $i=n-1$ until $i=1$.
So $(S_q,x)>0$ for all generators $L_x$ of $W_j$ because for
$i>0$ $L_x$ have common zero column.

With decreasing of $(S_q,x)$ and increasing $|R(x)|$, we can add to the set of nonzero columns 
 of matrices $L_x$ new columns due to $R(y)\supset R(x)$ and
$|R(y)|=|R(x)|+(S_q,x)-(S_q,y)$ (Lemma \ref{l7}).

The set of nonzero columns in matrices $W_j$ is a union of  nonzero columns of matrices from $V_i$ and 
in view of $(S_q,x)>0$ there is common zero column in matrices $W_j$.

One can extend the rank $|R(x)|$ and reduce $(S_q,x)$ of minimal solution $L_x$ of (\ref{ux}) only by decreasing
$|R(u)|$ (Lemma \ref{l7}, Corollary \ref{c7}).

We follow conditions of Lemma \ref{l7} with a view to obtain $(S_q,x)=0$ for solution $L_x$ of (\ref{ux}).
It's unavoidable after $\dim(W_j)>(n-1)^2$ (or before) in view of Corollary \ref{c2} and Lemma \ref{v8}.

The distinct linear independent solutions can be added consistently
extending the dimension of $W_j$ and upper bound $j$
of the length of the word $u$. So
\begin{equation}
\dim(W_j)=j+1 \quad |u|\leq j.     \label{d}
\end{equation}

\section{Theorems}

\begin{thm} \label{t}

The deterministic complete $n$-state synchronizing automaton
$A$ with strongly connected underlying graph over alphabet $\Sigma$ has synchronizing word in $\Sigma$
of length at most $(n-1)^2$.

\end{thm}

Proof.
The introduction to the former section considers a growing sequence of spaces $W_j$
(an ascending chain by dimension $j+1$) generated by linear
independent solutions $L_x$ of the equations (\ref{ux})
for $|u|\leq j$ by help of Lemmas \ref{l5} and \ref{l7} with Corollaries.

By Lemma \ref{v8}, any space $W$ generated by solutions $L_x$ of the equation $M_uL_x=M_s$ 
with common zero column and of restricted
length $|u|\leq j$ has a solution $L_x \not\in W$ for some word $u$ of length at most $j+1$.

 $\dim(W_j)\leq n(n-2)+1=(n-1)^2$ for $W_j$ with matrices
having units in at most $n-1$ column by Corollary \ref{c2}
of Lemma \ref{v3}.

So inevitably at least one solution $L_y\not\in W_j$ of equation (\ref{yx}) has corresponding word $v$ 
with $|v|=j+1>(n-1)^2$ and minimal $(S_q,y)=0$.

By Lemma \ref{l5}, for $L_x$ with minimal $(S_q,x)$ of equation (\ref{ux}) $|R(u)|-1=(S_q,x)$.
We reach finally a minimal $(S_q,y)=0$
for path of length $|v| \leq n(n-2)+1$.
So

\centerline{$|v|\leq n(n-2)+1$ with $|R(v)|=1$ and
$(S_q,y)=0$.}
Consequently the matrix $M_{v}$ of rank one in equation $M_vL_y=M_s$ is the matrix
of synchronizing word $v$ of length at most
$n(n-2)+1=(n-1)^2$.

\begin{cor} \label{c14}
For every integer $k<n$ of deterministic complete $n$-state synchronizing automaton $A$ with strongly connected
underlying graph over alphabet $\Sigma$
there exists a word $v$ of length at most $n(k-1)+1$ such that $|Av|\leq n-k$.
\end{cor}

\begin{cor} \label{c15}
For every set $P$ of states from deterministic complete $n$-state synchronizing automaton over alphabet $\Sigma$
there exists a word $s$ of length at most $(n-1)^2$ such that $|Ps|=1$.
\end{cor}

\begin{cor} \label{c16}
The graph $\Gamma^2$ of pairs of states for deterministic complete $n$-state synchronizing automaton
with underlying graph $\Gamma$ and $\Gamma s =\bf q$ for minimal  word $s$ has a set of paths to the pair (${\bf q}, \bf q$)
of length at most $(n-1)^2$ defined by $s$.

Every pair of distinct states belongs to a path from the set.
\end{cor}

\begin{thm}\label{t2}
The deterministic complete $n$-state synchronizing automaton
$A$ with underlying graph
over alphabet $\Sigma$ has synchronizing word in $\Sigma$ of length at most $(n-1)^2$.
\end{thm}
Follows from Theorem \ref{t} because the restriction for strongly connected graphs can be omitted due to \cite{Ce}.

\begin{thm}\label{t4}
Suppose that $|\Gamma\alpha|<|\Gamma|-1$ for a letter $\alpha \in\Sigma$ in deterministic complete $n$-state
synchronizing automaton $A$ with underlying graph $\Gamma$ over alphabet $\Sigma$.

Then the minimal length of synchronizing word of the automaton is less than $(n-1)^2$.

\end{thm}
Proof.
We follow the proof of Theorem  \ref{t}.

The difference is that at the beginning of the proof the equation
(\ref{ux}) has at least two linear independent nontrivial solutions
for the matrix $M_{\alpha}$ of a letter $\alpha$ equal to the first word $u$ of length one.

Hence we obtain finally synchronizing word of length less
than $(n-1)^2$.

Let us go to the case of not strongly connected underlying graph with $n-|I|>0$ states
outside minimal strongly connected ideal $I$.

This ideal has synchronizing word of length at most $(|I|-1)^2$ (Theorem \ref{t}).
There is a word $p$ of length at most $(n-|I|)(n-|I|+1)/2$ such that $Ap \subset I$.

$(|I|-1)^2 +(n-|I|)((n-|I|)+1)/2<(n-1)^2$.
Thus, the restriction for strongly connected automata can be omitted.

\begin{thm}\label{t5}
Every road coloring of edges of $n-state$ strongly connected directed graph
with constant outdegree and gcd=1 of length of all its cycles has synchronizing word
of length at most $(n-1)^2$.
\end{thm}

Proof follows from Theorem \ref{t} and work \cite{TP}.

\begin{thm}\label{t6}
Suppose that $|\Gamma\alpha|<|\Gamma|-1$ for a letter $\alpha \in\Sigma$ in deterministic complete $n$-state
synchronizing automaton $A$ with underlying graph $\Gamma$ over alphabet $\Sigma$.

Then the minimal length of synchronizing word of the automaton is less than $(n-1)^2$.
\end{thm}

\begin{proof}
We follow the proof of Theorem  \ref{t}.

The difference is that at the beginning of the proof there are at least two linear independent matrices
($L_{\alpha}$ and $L_v$ such that
$L_{\alpha}\sqsubseteq_q L_v$ for the letter $\alpha=u$.

So for the first word $u$ of length one there are at least two linear independent matrices with $\dim(W_1)>2$, i.e.
$\dim(W_j)>j$ for $j=1$.

Hence following the proof of Lemma \ref{v8} and Theorem \ref{t} we obtain synchronizing word of length less than $(n-1)^2$.

Let us go to the case of not strongly connected underlying graph
with $n-|I|>0$ states
outside minimal strongly connected ideal $I$.

This ideal has synchronizing word of length at most $(|I|-1)^2$ (Theorem \ref{t}).
There is a word $p$ of length at most $(n-|I|)(n-|I|+1)/2$ such that $Ap \subset I$.

$(|I|-1)^2 +(n-|I|)(n-|I|+1)/2<(n-1)^2$.
Thus the restriction for strongly connected automata can be omitted.
\end{proof}

\section{Examples}
J. Kari \cite{Ka} discovered the following example of $n$-state automaton
with minimal synchronizing word of length $(n-1)^2$ for $n=6$.

\begin{picture}(130,70)
 \end{picture}
\begin{picture}(130,74)
\multiput(6,60)(64,0){2}{\circle{6}}
\multiput(6,13)(64,0){2}{\circle{6}}
 \multiput(22,56)(22,0){2}{a}
\multiput(16,19)(34,0){2}{a}
 \put(36,21){\circle{6}}
\put(36,48){\circle{6}}
 \put(7,14){\vector(4,1){28}}
\put(7,57){\vector(4,-1){26}}

\put(39,52){\vector(4,1){27}}
 \put(37,20){\vector(4,-1){28}}
\put(67,63){\vector(-1,0){57}}
 \put(36,64){a}
\put(67,12){\vector(-1,0){57}}
 \put(32,0){a}

\put(70,15){\vector(0,1){42}}
 \put(70,59){\vector(0,-1){42}}
\put(34,21){\vector(1,1){36}}
 \put(52,28){b}

  \put(76,22){b}
\put(76,10){3}
\put(76,60){0}
\put(27,25){5}
\put(43,38){2}

\put(25,37){b}
\put(36,48){\circle{10}}
\put(-6,10){4}
\put(0,20){b}
\put(-6,60){1}
\put(0,45){b}
\put(37,42){\vector(2,1){4}}

\put(6,60){\circle{10}}
\put(4,64){\vector(2,1){4}}

 \put(6,13){\circle{10}}
\put(7,7){\vector(2,1){4}}
 \end{picture}

The minimal synchronizing word

\centerline{$s=\it ba^2bababa^2b^2aba^2ba^2baba^2b$}
has the length at the \v{C}erny border.

Every line below presents a pair (word $u$, $n$-vector $c_u$) of linear independent matrices $L_u$ from the sequence.

$(b, 111110)$ $R(u)=5$

$(ba, 111011)$

$(ba^2, 111101)$

$(ba^2b, 111100)$ $R(u)=4$

$(ba^2ba, 111010)$

$(ba^2bab, 011110)$

$(ba^2baba, 101111 )$ $R(v)=5$ (l01011 of $L_u$)

$(ba^2babab, 101110)$ $R(u)=4$

$(ba^2bababa, 110101)$

$(ba^2bababa^2, 011101)$

$(ba^2bababa^2b, 111000)$ $R(u)=3$

$(ba^2bababa^2b^2, 011100)$

$(ba^2bababa^2b^2a, 110111)$ $R(v)=5$ (101010 of $L_u$)

$(ba^2bababa^2b^2ab, 001110)$ $R(u)=3$

$(ba^2bababa^2b^2aba, 100011)$

$(ba^2bababa^2b^2aba^2, 011111)$  $R(v)=5$ (010101 of $L_u$)

$(ba^2bababa^2b^2aba^2b, 110000)$ $R(u)=2$

$(ba^2bababa^2b^2aba^2ba, 011000)$

$(ba^2bababa^2b^2aba^2ba^2, 101000)$

$(ba^2bababa^2b^2aba^2ba^2b, 001101)$ $R(v)=3$ (001100 of $L_u$)

$(ba^2bababa^2b^2aba^2ba^2ba, 100010)$ $R(u)=2$

$(ba^2bababa^2b^2aba^2ba^2bab, 000110)$

$(ba^2bababa^2b^2aba^2ba^2baba, 001011)$  $R(v)=3$ (000011 of $L_u$)

$(ba^2bababa^2b^2aba^2ba^2baba^2, 000101)$ $R(u)=2$

$(ba^2bababa^2b^2aba^2ba^2baba^2b=s, 100000)$  $R(s)=1$

By the bye, the matrices of right subwords of $s$ are simply linear independent.
\\
\\
For the \v{C}erny sequence of $n$-state automata \cite{Ce}, \cite{La}, \cite{Li} the situation is more pure.

\begin{picture}(300,70)
\multiput(0,54)(26,0){14}{\circle{6}}
\multiput(26,54)(26,0){13}{\circle{10}}
\multiput(24,54)(26,0){6}{\vector(-1,0){20}}
\multiput(204,54)(26,0){6}{\vector(-1,0){20}}
\put(160,54){ ....}

\multiput(11,58)(26,0){13}{a}
 \multiput(26,64)(26,0){13}{b}
 \multiput(25,58)(26,0){13}{\vector(2,1){4}}
\put(340,17){\vector(0,1){34}}
 \put(0,51){\vector(0,-1){34}}
\put(2,51){\vector(0,-1){34}}
\put(-7,34){a} \put(6,34){b} \put(330,34){a}
\multiput(0,13)(26,0){14}{\circle{6}}
\multiput(0,13)(26,0){14}{\circle{10}}
\multiput(4,13)(26,0){6}{\vector(1,0){20}}
\multiput(186,13)(26,0){6}{\vector(1,0){20}}
\put(160,13){ ....}

\multiput(12,15)(26,0){13}{a}
 \multiput(0,-3)(26,0){14}{b}
 \multiput(-3,17)(26,0){14}{\vector(2,1){4}}
 \end{picture}
\\
\\
The minimal synchronizing word

\centerline{$s=b(a^{n-1}b)^{n-2}$}
 of the automaton also has the length at the \v{C}erny border.
\\
\\
For $n=4$

\begin{picture}(140,54)
\end{picture}
\begin{picture}(140,50)
\multiput(-21,60)(60,0){2}{\circle{6}}
\multiput(-21,60)(60,0){2}{\circle{10}}

 \multiput(-21,10)(60,0){2}{\circle{6}}

\put(39,10){\circle{10}}

 \put(-21,12){\vector(0,1){42}}
\put(-19,12){\vector(0,1){42}}

 \put(34,10){\vector(-1,0){51}}
\put(-16,60){\vector(1,0){50}}
 \put(39,55){\vector(0,-1){40}}

  \multiput(-29,35)(60,0){2}{a}
\multiput(10,5)(0,60){2}{a}

\put(-20,67){2 b}
 \put(-17,-1){1}
\put(38,67){3  b}
 \put(38,-3){4  b}
\put(-18,35){b}

 \end{picture}

and synchronizing word  $baaabaaab$ with pairs of word $u$ and
$n$-vector $c_u$ of linear independent matrices $L_u$ below.

$(b, 0111)$ $R(u)=3$

$(ba, 1011)$

$(baa, 1101)$

$(baaa, 1110)$

$(baaba, 1010)$  $R(u)=2$

$(baaaba, 0011)$

$(baaabaa, 1001)$

$(baaabaaa, 1100)$ $k=8$

$(baaabaaab=s, 0100)$ $R(s)=1$
\\
\\
In the example of Roman \cite{Ro}

\begin{picture}(140,54)
\end{picture}
\begin{picture}(140,50)
\multiput(-21,39)(56,0){3}{\circle{6}}
\multiput(-21,39)(56,0){3}{\circle{10}}

 \multiput(6,10)(60,0){2}{\circle{6}}

\put(66,10){\circle{10}}
 \put(5,12){\vector(-1,1){24}}
  \put(-19,36){\vector(1,-1){24}}

\put(67,12){\vector(1,1){22}}
  \put(91,36){\vector(-1,-1){22}}

  \multiput(-16,20)(97,0){2}{c}

\put(7,12){\vector(1,1){24}}
\put(31,36){\vector(-1,-1){24}}

\put(38,36){\vector(1,-1){24}}

   \put(9,10){\vector(1,0){54}}
   \put(63,10){\vector(-1,0){54}}

\put(2,-1){3}
 \put(28,0){a}
\put(53,25){a}
 \put(11,24){b}

\put(-33,41){$5$}
\put(-14,45){$a,b$}
\put(28,48){$c$}
\put(19,41){$2$}
\put(98,45){$a,b$}
\put(88,47){$4$}
\put(60,-2){b   1}
 \end{picture}

the minimal synchronizing word

\centerline{$s=ab(ca)^2c$ $bca^2c$ $abca$}

The line below presents a pair (word $u$, $n$-vector $c_u$) of linear independent matrices $L_u$.

$(a, 10111)$ $R(u)=4$

$(ab, 11011)$

$(abc, 11110)$

$(abca, 10110)$ $R(u)=3$

$(abcac, 10011)$

$(abcaca, 01111)$ $R(v)=4$  (00111 of $L_u$)

$(abcacac, 10101)$ $R(u)=3$

$(abcacacb, 11001)$

$(abcacacbc, 01110)$

$(abcacacbca, 10010)$ $R(u)=2$

$(abcacacbca^2, 00110)$

$(abcacacbca^2c, 10001)$

$(abcacacbca^2ca, 11101)$  $R(v)=4$ (00101 of $L_u$)

$(abcacacbca^2cab, 01001)$   $R(u)=2$

$(abcacacbca^2cabc, 01100)$

$(abcacacbca^2cabca=s, 10000)$ $R(s)=1$

\section*{Acknowledgments}
I would like to express my gratitude to Francois Gonze, Dominique Perrin, Marie B{\'e}al, Akihiro Munemasa, Wit Forys, Benjamin Weiss, Mikhail Volkov and Mikhail Berlinkov for fruitful and essential remarks.


\begin{thebibliography}{99}
\bibitem{AW} R.L. Adler, B. Weiss:  Similarity of automorphisms
of the torus, Memoirs of the Amer. Math. Soc., 98, Providence, RI,
 1970.
\bibitem{AGV} D. S. Ananichev, V. V. Gusev, and M. V. Volkov.
Slowly synchronizing automata and digraphs. Springer, Lect. Notes in Comp. Sci., 6281(2010), 55-65.
\bibitem{Be} M.-P. B{\'e}al, A note on \v{C}erny Conjecture and rational series,
 technical report, Inst. Gaspard Monge, Univ. de Marne-la-Vallee, 2003.
\bibitem{BBP} M.-P. B{\'e}al, M.V. Berlinkov, D. Perrin. A quadratic upper bound on
the size  of a synchronizing word in one-cluster automata, Int. J. Found. Comput. Sci. 22(2), 2011, 277-288.
\bibitem{BR} J. Berstel, C. Reutenauer, Rational series and their languages, Springer, 1988.
\bibitem{CA} A. Carpi, F. D'Alessandro, Strongly transitive automata and the  \v{C}erny conjecture. Acta Informatica, 46(2009), 591-607.
\bibitem{Ce} J. \v{C}erny, Poznamka k homogenym eksperimentom s konechnymi automatami, Math.-Fyz. \v{C}as., 14(1964), 208-215.
\bibitem{Co} A. Carpi, F. D'Alessandro, On the Hybrid \v{C}erny-Road coloring problem
and Hamiltonian paths. LNCS, 6224(2010), 124-135.
\bibitem{CR} K.Chmiel, A. Roman. COMPAS - A Computing Package for Synchronization.
LNCS, Impl. and Appl. of Automata, 6482(2011), 79-86, 2011.
\bibitem{CPR} J. \v{C}erny, A. Piricka, B. Rosenauerova.
On directable automata, Kybernetika 7(1971), 289-298.
\bibitem{DZ} H. Don, H. Zantema, Finding DFAs with maximal shortest synchronizing word length. arXiv:1609.06853, 2016.
 \bibitem{Fr} P. Frankl, An extremal problem for two families of sets, Eur. J. Comb., 3(1982), 125-127.
\bibitem{Fi} J. Friedman. On the road coloring problem, Proc. of the Amer. Math. Soc. 110(1990), 1133-1135.
 \bibitem{GJT} F. Gonze, R. M. Jungers, A.N. Trahtman.  A Note on a Recent Attempt to Improve the Pin-Frankl Bound. DM \& TCS, 1(17), 2015, 307-308.
\bibitem{Ju} H. Jurgensen, Synchronization. Inf. and Comp. 206(2008), 9-10, 1033-1044.
\bibitem{Ka} J. Kari, A counter example to a conjecture concerning synchronizing word in finite automata, EATCS Bulletin, 73(2001), 146-147.
\bibitem{Kr} J. Kari,  Synchronizing finite automata on Eulerian digraphs. Springer, LNCS, 2136(2001), 432-438.
\bibitem{KV} J. Kari, M. V. Volkov, \v{C}erny's conjecture and the road coloring problem. Handbook of Automata, 2013.
\bibitem{KKS}  A. Kisielewicz, J. Kowalski, M. Szykula. Computing the shortest reset words of synchronizing automata, J. Comb. Optim., Springer, 29(2015), 88-124.
\bibitem{KRS}  A.A. Kljachko, I.K. Rystsov, M.A. Spivak, An extremely
combinatorial problem connected with the bound on the length of a recurrent
word in an automata. Kybernetika. 2(1987), 16-25.
\bibitem{Ln} P. Lankaster, Theory of matrices, Acad. Press, 1969.
\bibitem{La} A. E. Laemmel. Study on application of coding theory.
Technical Report PIBMRI-895. Dept. Electrophysics, Microwave Research Inst., Polytechnic Inst., Brooklyn, NY, 5-63, 1963.
\bibitem{Li} C. L. Liu. Determination of the final state of an automaton whose initial state is unknown.
IEEE Transactions on Electronic Computers, EC-12(5):918-920, 1963.
 \bibitem{Ma} A. I. Malcev, Foundations of linear algebra,
San Francisco, Freeman, 1963. (Nauka, 1970, in Russian.)
\bibitem{MS} A. Mateescu and A. Salomaa, Many-Valued Truth Functions, Cerny's conjecture and road coloring, Bulletin EATCS,  68 (1999), 134-148.
\bibitem{PS} D.  Perrin,  M.-P. Schutzenberger. Synchronizing prefix codes and automata
and the road coloring problem. Symbolic dynamics and its applications , 135 (1992),
295-318.
\bibitem{Pin} J.-E. Pin, On two combinatorial problems arising
 from automata theory, Annals of Discrete Math., 17(1983), 535-548.
\bibitem{Ro} A. Roman, Experiments on Synchronizing Automata.
Schedae Informaticae, Versita, Warsaw, 19(2010), 35-51.
\bibitem{Sta} P. H. Starke. Eine Bemerkung ueber homogene Experimente. Elektronische Informationverarbeitung und Kybernetik, 2(1966), 257-259.
\bibitem{St} B.  Steinberg, The Averaging Trick and the Cerny Conjecture.
Developments in Language Theory, Springer, NY, LNCS, 6224(2010), 423-431.
\bibitem{Sz} M. Szykula. Improving the Upper Bound on the Length of the Shortest Reset Words. STACS 2018 v. 96, 56:1--56:13.
\bibitem{TS} A.N. Trahtman.  Notable trends concerning the synchronization
of graphs and automata, CTW06, El. Notes in Discrete Math., 25(2006), 173-175.
\bibitem{Ta} A.N. Trahtman.  The \v{C}erny Conjecture for Aperiodic Automata.
 Discr. Math. Theoret. Comput. Sci. v. 9, 2(2007), 3-10.
\bibitem{TP} A.N. Trahtman.  The Road Coloring and Cerny Conjecture.
Proc. of Prague Stringology Conference.  2008, 1-12.
\bibitem{TF} A.N. Trahtman. Modifying the upper bound on the length of minimal synchronizing word. LNCS, 6914(2011), 173-180, arXiv:1104.2409.
\bibitem{TB} A.N. Trahtman. Bibliography, http://u.cs.biu.ac.il/$\sim$trakht/syn.html.
\bibitem{Tt} A.N. Trahtman. Synchronization, http://u.cs.biu.ac.il/$\sim$trakht/readme.html.
\bibitem{Vo} M. V. Volkov, Synchronizing automata and the Cerny conjecture, in: C.Martin-Vide, F. Otto, H. Fernau eds., Language and Automata Theory and Applications, LATA 2008, Springer, LNCS, 5196(2008), 11-27.
\bibitem{Vo1} M. V. Volkov, Lecture Notes on Synchronizing Automata,  M, V. Volkov, 2016,http://csseminar.kadm.usu.ru/Hunter/lectures.pdf.
\end{thebibliography}
 \end{document}